\documentclass[conference]{IEEEtran}
\IEEEoverridecommandlockouts
\def\BibTeX{{\rm B\kern-.05em{\sc i\kern-.025em b}\kern-.08em
    T\kern-.1667em\lower.7ex\hbox{E}\kern-.125emX}}
    
\usepackage{cite}
\usepackage{amsmath,amssymb,amsfonts}
\usepackage{algorithmic}
\usepackage{graphicx}
\usepackage{textcomp}
\usepackage{xcolor}
\usepackage{float}
\usepackage{multicol}
\usepackage{verbatim}
\usepackage{amsmath,amssymb}
\usepackage{algorithm}
\usepackage{float} 
\usepackage{booktabs} % For formal tables
\usepackage{tabularx} % For adjustable-width columns
\usepackage{caption}  % For custom captions
\usepackage{graphicx} % For including images
\usepackage{subcaption} % For subfigures

\begin{document}

\title{Global Scale Self-Supervised Channel Charting
with Sensor Fusion
}

\author{
\IEEEauthorblockN{Omid Esrafilian$^{(1)\,*}$, Mohsen Ahadi$^{(1)\,*}$, Florian Kaltenberger$^{(1) (2)}$, David Gesbert$^{(1)}$}
\IEEEauthorblockA{$^{(1)}$Communication Systems Department, EURECOM, Sophia Antipolis, France\\
$^{(2)}$Institute for the Wireless Internet of Things, Northeastern University, Boston, USA\\
Emails: \{omid.esrafilian, mohsen.ahadi, florian.kaltenberger, david.gesbert\}@eurecom.fr
% , f.kaltenberger@northeastern.edu
}
\thanks{$^{*}$ Equal contributions.
}
}

\maketitle

\begin{abstract}
The sensing and positioning capabilities foreseen in 6G have great potential for technology advancements in various domains, such as future smart cities and industrial use cases. Channel charting has emerged as a promising technology in recent years for radio frequency-based sensing and localization. However, the accuracy of these techniques is yet far behind the numbers envisioned in 6G. To reduce this gap, in this paper, we propose a novel channel charting technique capitalizing on the time of arrival measurements from surrounding Transmission Reception Points (TRPs) along with their locations and leveraging sensor fusion in channel charting by incorporating laser scanner data during the training phase of our algorithm. The proposed algorithm remains self-supervised during training and test phases, requiring no geometrical models or user position ground truth. Simulation results validate the achievement of a sub-meter level localization accuracy using our algorithm 90$\%$ of the time, outperforming the state-of-the-art channel charting techniques and the traditional triangulation-based approaches.
\end{abstract}

% \begin{IEEEkeywords}

% \end{IEEEkeywords}

\section{Introduction}
%Indoor User Equipment (UE) positioning in 5G New Radio %(NR) represents a significant advancement in wireless %technology, offering highly accurate location services %within indoor environments. This technology leverages %the enhanced bandwidth and lower latency of 5G networks %to provide precise and reliable location %tracking\cite{ETSI_TS_138_305}. Key use cases include %localizing tools in smart factories, asset tracking in %warehouses, healthcare monitoring, etc.
%However, the complexity of such indoor environments %significantly impacts the effectiveness and reliability %of positioning and communication systems. One of the %key challenges in such environments is the high %probability of Non-Line-of-Sight (NLoS) conditions.\\
Positioning techniques in wireless networks traditionally rely on channel parameter estimation methods like Received Signal Strength Indicator (RSSI), Time of Arrival (ToA), Time Difference of Arrival (TDoA), and Angle of Arrival/Departure (AoA/AoD) \cite{3gpp38.305,10119056}. These parameters help triangulate or trilaterate a device's position while Channel State Information (CSI) offers a more comprehensive approach by considering detailed channel properties and environmental effects on the signal. This method is particularly effective in complex indoor environments where dense multipath propagation and NLoS conditions are dominant.
Direct CSI positioning methods are divided into two categories, supervised and unsupervised\cite{9264122}.
Supervised learning, particularly fingerprinting (FP) methods, relies on a pre-established database of signal characteristics like CSI or Multi Path Components (MPC) as features collected at known UE locations. This database is used to train a model that can later predict the location of a device based on its observed CSI or features, making it highly effective in environments with stable signal patterns. Unsupervised learning, on the other hand, does not require a pre-labeled dataset. Instead, it estimates a mapping directly from the collected data for an objective function. This approach is valuable in dynamic environments where it is impractical to create and maintain an extensive labeled dataset, offering the flexibility to adapt to changes in the environment over time.\\
Although CSI provides a comprehensive view of wireless signal propagation from the transmitter to the receiver, including the effects of environmental factors (e.g., scattering, fading, and reflection), this detailed information results in high-dimensional data, making analysis and application in positioning systems challenging. Channel Charting (CC), an innovative approach in wireless networking, seeks to create a map or chart of the wireless medium using CSI, enabling the precise localization and tracking of devices within complex environments \cite{ferrand2023wireless}. Channel charting is one of the applications of manifold learning, a non-linear dimensionality reduction technique, which plays a vital role in interpreting CSI data. Manifold learning effectively uncovers the low-dimensional structures hidden within high-dimensional CSI datasets. This refines signal propagation properties, resulting in more precise and efficient perceptions of wireless signal interactions with their surroundings. The primary objective in dimensionality reduction is to map data from a high-dimensional space (D) to a lower-dimensional space (d) (where $d << D$), with two key objectives: i) The mapping should maintain the proximal relationships among data points, and ii) It should effectively generalize to new and unseen data.
Non-parametric CC techniques such as Multidimensional Scaling (MDS), Isometric Mapping (ISOMAP), and Principal Component Analysis (PCA) perform well in simplifying high-dimensional data while retaining critical structures \cite{van2009dimensionality}. MDS focuses on preserving pairwise distances, ISOMAP extends this by maintaining geodesic distances on a manifold, and PCA captures the maximum variance through orthogonal principal components. However, these methods often fail to accurately predict for unseen data as they do not learn a general mapping function from the original high-dimensional space to the reduced space. Instead, they are entirely dependent on the specific dataset they are applied to, and any new data requires recalculating the entire model, making them unsuitable for prediction in dynamic environments.
Conversely, parametric approaches like deep learning-based methods  are adept at learning a mapping function that can be applied to novel, unseen data. This capacity to extend CC to previously unseen CSI data is a significant advantage over traditional strategies.
Despite the recent advancements in self-supervised CC using deep metric learning, they still struggle to match the precision achieved by supervised or traditional triangulation methods even in the Line-of-sight (LoS) state. Consequently, in this paper, we developed a novel channel charting algorithm, leveraging neural network and data fusion to accurately localize the user. Specifically, our contributions are as follows:

% Despite the recent advancements in self-supervised CC using deep metric learning, they still struggle to match the precision achieved by supervised or traditional triangulation methods even in the LoS state. Consequently, we leverage sensor fusion to enhance self-supervised positioning using integrated CSI and laser data only in the training phase. This enables our approach to preserve the global geometry features, resulting in an accurate localization with a sub-meter error in $\text{90}\%$ of the test cases using at least $2$ LoS Transmission Reception Points (TRPs).

\begin{itemize}
\item A neural network-based channel charting function accurately localizes users while preserving global geometry.
\item We enhanced localization accuracy using data fusion with depth data during training.
\item Our algorithm is self-supervised, utilizing nearby Transmission Reception Points (TRPs) locations and depth data during training without requiring labeled data.
\item Our method achieves a sub-meter localization accuracy using two LoS TRPs in 90$\%$ of the time, superior to the state-of-the-art and traditional triangulation methods.
\end{itemize}

\section{Related work}
Channel charting for localization in wireless networks has been used for the first time in \cite{8444621} from a single base station (BS) with multiple antennas, and in \cite{8645281,9109875} from multiple massive MIMO BSs in space. Since CC relies on dimensionality reduction of the CSI data, \cite{Huang2019ImprovingCC} and \cite{9771913} used autoencoders to improve this task. A Siamese neural network in \cite{8919897,10070385} is proposed that takes random pairs of CSI data to first learn a local channel chart and then transform it to the global form using a subset of labeled data as reference points in a semi-supervised manner. In this method, the Euclidean distances of the CIR measurements are used as a dissimilarity metric. To overcome the limitations of the Siamese loss function with a Euclidean distance metric, a triplet-based loss is used in \cite{9448128,9833925} to learn the similarity between triplets of CSI data based on the distance of other side-informations such as the relative recording timestamps. Authors of 
\cite{10074200,euchner2023augmenting} combined CC with the classical localization approaches, taking ToA and AoA measurements to improve the global channel chart. Although the CSI measurements can contain rich information, none of these CC studies exploiting only CSI data have surpassed the performance of traditional triangulation-based methods, even when LoS conditions are present. In \cite{stahlke2023velocitybased}, velocity estimation and topological map data are used for the global transformation of the CC. However, the global consistency of this algorithm relies on the length of the trajectory taken by the user. Also, the map-matching algorithm in this study works only if a unique match of the channel chart exists in the map.
Finally in \cite{taner2023channel}, by proposing a loss function containing a bilateration loss including multiple BSs with known locations and a triplet loss, a self-supervised CC is made in real-world coordinates. Motivated by this, in this paper, we leverage sensor fusion and the location of the nearby TRPs in the training phase to enhance localization accuracy. 
% Ultimately, without maintaining a supervised approach in their training or chart transformation, none of the CC studies have surpassed the performance of traditional triangulation-based methods, even when LoS conditions are present.

\section{System Model}
We consider an indoor scenario where a mobile user, a.k.a UE, \footnote{By no means, the algorithm proposed in this paper is limited to one user and can be generalized to the multiple UE scenario.} communicates with $M$ fixed TRPs with known locations $\textbf{x}_m\in \mathbb{R}^3, \forall m\in[1, M]$ placed in the environment. The UE follows a trajectory with a duration of $N$ time steps. The UE location at each time step $n$ is assumed unknown and denoted by $\textbf{u}_n \in \mathbb{R}^3, n\in [1, N]$. The height of the UE is presumed to be fixed during its trajectory. Both the UE and TRPs are equipped with single antennas and operate in orthogonal frequency-division multiplexing (OFDM) transmission mode with total $C$ sub-carriers. The estimated CSI of the link between the $m$-th TRP and the UE location $\textbf{u}_n$ (yet assumed unknown) at time step $n$ over all sub-carriers is denoted by $\textbf{h}_{m, n} \in \mathbb{C}^C$. We assume that the UE and each TRP are synchronized.
%, however, a tight synchronization between TRPs is not required%
We denote the channel impulse response (CIR) by applying an inverse discrete Fourier transform on each CSI vector as $\textbf{w}_{m, n}$ and the concatenated CIR matrix over all TRPs at time step $n$ is given by $\textbf{W}_{n} \in\mathbb{C}^{M\times C}$. We also consider two phases of train and testing in this paper. During the training phase, the UE is equipped with a 2D laser scanner to collect depth measurements in addition to CSI, while during the test time, the algorithm only uses the CSI measurements, and no laser scanner is required. It is worth mentioning that, the algorithm remains self-supervised during both train and test phases and the only difference is the type of measurements available at each phase. The depth measurements collected by the laser scanner at time step $n$ during the training phase are denoted by $\boldsymbol{\ell}_n = \{ (r_k, \phi_k), \forall k\in[1, K]\}\in \mathbb{R}^{K\times 2}$, where $K$ is the number of points at each scan of the laser scanner, $r_k,$ and $\phi_k$ are the relative distance and the angle of each scanned point, respectively, in the laser scanner body coordinate frame which we assume to be the same as the UE coordinate frame.

\subsection{Data Preprocessing  and Feature Extraction}
To increase the robustness of the algorithm, we preprocess and extract certain features from the measured CIR.  Since the majority of the received power is usually concentrated in the first few taps, we only consider the first $\bar{C}$ columns of the CIR matrix. 
% To further reduce the sensitivity to the transmit power, we normalize the CIR. 
The truncated CIR is denoted by $\hat{\textbf{W}}_{n} \in \mathbb{C}^{M\times \bar{C}}$. Therefore, the main input of our algorithm is computed as $\textbf{Y}_n = | \hat{\textbf{W}}_{n}|\in \mathbb{R}^{M\times \bar{C}}$, where $|.|$ is the element-wise absolute value operator. We also extract the ToA of the LoS path between UE and TRPs at time step $n$ by detecting the largest pick among the columns of  $\textbf{Y}_n$. Equivalently, we can write:
\begin{equation}
    \tau_{m, n} \propto \arg \max_{c\in[1, \bar{C}]} y_n^{(m)(c)},
\end{equation}
\noindent where $ y_n^{(m)(c)}$ is the element of matrix $\textbf{Y}_n$ at row $m$ and column $c$, and $\tau_{m, n}$ is the measured ToA corresponding to the LoS path (i.e. the largest pick of the CIR vector) between the UE and the $m$-th TRP at time step $n$. The ToA of all TRPs and the UE at time step $n$ in a vectorized form is given by $\boldsymbol{\tau}_n = (\tau_{1, n}, \cdots, \tau_{M, n}) \in \mathbb{R}^M$. 

Furthermore, we exploit the additional measurements provided by a 2D laser scanner, which is available only during the training phase, to estimate the UE displacement between two different time steps. To this end, we employ an Iterative Closest Point (ICP) algorithm which is a well-known algorithm for processing the laser scanner data. For the sake of limited space, we omit the details of the ICP algorithm and we refer to \cite{besl1992method} for more information. We denote the estimated UE displacement between time steps $n$ and $n'$ using ICP algorithm and utilizing laser scanner data by $\hat{T}_{n, n'}$.

\subsection{Deep Channel Charting}
Given the CIR dataset, it is possible to find a mapping function $f_{\boldsymbol{\theta}}:\mathbb{R}^{M\times \bar{C}} \rightarrow  \mathbb{R}^{D}$ that transforms the CIR matrix $\textbf{Y}_n$  to a lower dimension $D \leq 3$ as a proxy to the locations, a.k.a pseudo-position, of the user as $\Tilde{\textbf{u}}_n = f_{\boldsymbol{\theta}}(\textbf{Y}_n)$. Deep neural network has proven to be a good candidate for estimating the mapping function $f_{\boldsymbol{\theta}}$, as $f_{\boldsymbol{\theta}}$ is a complicated and non-linear function. Various methods have been introduced in the literature to find the mapping function using deep neural networks. These methods range from supervised to unsupervised \cite{8444621,8645281,9109875,Huang2019ImprovingCC, 9771913, 8919897,10070385, 9448128,9833925, 10074200,euchner2023augmenting, stahlke2023velocitybased, taner2023channel}. 
% Some of the commonly used techniques include Siamese neural networks [], autoencoders[], and combined triplet and bilateration loss functions trained with data timestamps []. 
In this paper, we build a channel chart algorithm upon the bilateration loss function akin to \cite{taner2023channel} and by capitalizing on ToA measurements and the location of the TRPs. We extend this method further by incorporating laser scanner data to improve the accuracy of localization. Note that, the TRPs locations and laser scanner data are only required during the training phase. Moreover, our approach is self-unsupervised and will provide a global scale representation of the user's location in the global coordinate frame very close to the ground truth as opposed to the pseudo-position of the user. In the following section, we elaborate on our approach.

\section{Channel Charting Using Data Fusion} \label{sec:CCDateFusion}
In this section, we seek to learn a channel chart function $f_{\boldsymbol{\theta}}$ given a training dataset $\mathcal{D}_{\text{tr}} = \{\textbf{Y}_n, \boldsymbol{\tau_{n}}, \textbf{x}_m, \boldsymbol{\ell}_n; \forall n, m \}$. We assume that the pilot signal sent by the UE is received at all TRPs. As expressed in \cite{taner2023channel}, from the received CSI, when we compare the relative received powers at TRPs for a given UE, the TRPs closer to the UE tend to receive a signal with higher power under LoS conditions. Let's denote the received power at TRP $m$ from the UE at time step $n$ with $\gamma_{m, n} = 20 \log(\| \textbf{h}_{n, m} \|_F)$, where $\|. \|_F$ is the Frobenius norm. Therefore, we can write:
\begin{equation} \label{eq:far_close_power_diff}
    \gamma_{m, n_c} > \gamma_{m, n_f} + \Gamma, \forall n_c, n_f \in [1, N],
\end{equation}
where $n_c, n_f$ are time steps chosen such that the UE location at time step $n_c$ is closer to TRP $m$ than when the UE is at time step $n_f$. In other words, $n_c, n_f$ should satisfy the following:
\begin{equation} \label{eq: far_close_dist_diff}
    \| \textbf{x}_m - \textbf{u}_{n_c} \| < \| \textbf{x}_m - \textbf{u}_{n_f} \|,
\end{equation}
\noindent where $\|. \|$ is the Euclidean norm.
The constant $\Gamma$ imposes that the received power differs at least by $\Gamma$. 
% Assuming that the TRPs are LoS to the UE at all time steps and from the geometry, expressions in \eqref{eq: far_close_power_diff} and \eqref{eq: far_close_dist_diff} will lead to the following bilateral loss function:
From \eqref{eq: far_close_dist_diff}, we can constitute the following bilateration loss function:
\begin{equation} \label{eq:far_close_dist_loss}
\mathcal{L}^{b}_{m, n_c, n_f} = \max(\| \textbf{x}_m - f_{\boldsymbol{\theta}}(\textbf{Y}_{n_c}) \| - \| \textbf{x}_m - f_{\boldsymbol{\theta}}(\textbf{Y}_{n_f})\|  + d, 0),
\end{equation}
where $d > 0$ indicates that the UE at estimated location $f_{\boldsymbol{\theta}}(\textbf{Y}_{n_c})$ is closer to TRP $m$ than the UE estimated location $f_{\boldsymbol{\theta}}(\textbf{Y}_{n_f})$ by at least $d$ meters. Assuming that the TRPs are LoS to the UE at all time steps, finding a $f_{\boldsymbol{\theta}}$ that minimizes \eqref{eq:far_close_dist_loss} will ideally satisfy inequality \eqref{eq:far_close_power_diff}. However, in a realistic scenario, there is no guarantee that the power-distance relation holds, as also indicated in \cite{taner2023channel}. Moreover, the loss in \eqref{eq:far_close_dist_loss} might be equal to zero for a vast majority of $(n_c, n_f)$ pairs, depending on the value chosen for $d$, rendering it sample inefficient for learning $f_{\boldsymbol{\theta}}$. To tackle this problem, since we can estimate the ToA at each TRP, therefor we can have an estimate of $d$ per measurement at each time step. Consequently, we can reformulate the loss function in \eqref{eq:far_close_dist_loss} as follows
\begin{equation} \label{eq:far_close_dist_loss_2}
\mathcal{L}^{b}_{m, n_c, n_f} = (\| \textbf{x}_m - f_{\boldsymbol{\theta}}(\textbf{Y}_{n_c}) \| - \| \textbf{x}_m - f_{\boldsymbol{\theta}}(\textbf{Y}_{n_f})\|  + d_{m, n_c, n_f})^2,
\end{equation}
with
\begin{equation}
    d_{m, n_c, n_f} = |\tau_{m, n_c} - \tau_{m, n_c}| \, \nu,
\end{equation}
where $\nu$ is the speed of light, and $|.|$ represents the absolute value operator. Please note that, the loss function in \eqref{eq:far_close_dist_loss_2} is specified per each measurement pair and is not spars, resulting in a more sample-efficient training process. However, the expression in \eqref{eq:far_close_dist_loss_2} is a 
% highly nonlinear and non-convex function, hence challenging to solve with numerical gradient-based algorithms, such as gradient descent. 
differential loss function with respect to the UE location in two time steps and can bring ambiguity to the UE location estimate, hence not preserving the global geometry features.
To solve this issue, we redefine $\mathcal{L}^{b}_{m, n_c, n_f}$ by splitting it  into two additive parts as follows:
\begin{equation} \label{eq:far_close_dist_loss_3}
\mathcal{L}^{b}_{m, n_c, n_f} \triangleq \mathcal{L}_{m, n_c} + \mathcal{L}_{m, n_f},
\end{equation}
\noindent where 
\begin{equation}
     \mathcal{L}_{m, n} = (\| \textbf{x}_m - f_{\boldsymbol{\theta}}(\textbf{Y}_{n}) \|  - \tau_{m, n} \, \nu)^2.
\end{equation}
Furthermore, We can formulate a separate loss function for the UE locations at time steps $n_c, n_f$ as follows:
\begin{equation} \label{eq:displacement_loss}
\mathcal{L}^{\ell}_{n_c, n_f} = (\|f_{\boldsymbol{\theta}}(\textbf{Y}_{n_c})  - f_{\boldsymbol{\theta}}(\textbf{Y}_{n_f}) \| - \|\textbf{u}_{n_c} -\textbf{u}_{n_f} \|)^2,
\end{equation}
\noindent where the first part of this loss function $(\|f_{\boldsymbol{\theta}}(\textbf{Y}_{n_c})  - f_{\boldsymbol{\theta}}(\textbf{Y}_{n_f}) \|)$ represents the displacement between two estimated UE locations at time steps $n_c, n_f$, and the second part $ (\|\textbf{u}_{n_c} -\textbf{u}_{n_f} \|)$ is equivalent to the true UE displacement. Therefore, minimizing  \eqref{eq:displacement_loss} will result in preserving the relative displacement between the estimated UE locations in two different time steps. However, the second part of this loss function is not available. To tackle this problem, we exploit the measurements obtained from the laser scanner. We reformulate the loss function \eqref{eq:displacement_loss} by incorporating the laser scanner data as follows:
\begin{equation} \label{eq:laser_loss}
\mathcal{L}^{\ell}_{n_c, n_f} = (\|f_{\boldsymbol{\theta}}(\textbf{Y}_{n_c})  - f_{\boldsymbol{\theta}}(\textbf{Y}_{n_f}) \| - \hat{T}_{n_c,n_f})^2,
\end{equation}
\noindent where $\hat{T}_{n_c,n_f}$ is the estimated UE displacement between two time steps $n_c, n_f$ using laser scanner data and the ICP algorithm \cite{besl1992method}.

Finally, the total loss function including the CIR radio measurements and the laser scanner data is given by:
\begin{equation} \label{eq:total_loss}
\mathcal{L}_{m, n_c, n_f} = \mathcal{L}^{b}_{m, n_c, n_f} + \lambda_{n_c, n_f} \mathcal{L}^{\ell}_{n_c, n_f},
\end{equation}
\noindent where $\lambda_{n_c, n_f}$ is a coefficient that determines the impact of the loss pertaining to the laser scanner data. It is worth mentioning that the ICP algorithm might fail to estimate the displacement between two time steps if the depth measurements in two corresponding scans are very different. Therefore, to mitigate this error, we choose the value for $\lambda_{n_c, n_f}$ to be small when $n_c$ and $ n_f$ are far in time. 

Finally, the total loss function for all TRPs and over all time steps is given by:
\begin{equation}
    \mathcal{L} = \sum_{\substack{n_f, n_c \in [1, N] \\ n_f \neq n_c}}\sum_{m=1}^{M} \mathcal{L}_{m, n_c, n_f}.
\end{equation}
\noindent A neural network can then be trained to obtain a channel chart function $f_{\boldsymbol{\theta}}$ by minimizing $\mathcal{L}$ and using training dataset $\mathcal{D}_{\text{tr}}$. We denote the trained channel chart neural network model by $f_{\boldsymbol{\theta^*}}$, where $\theta^*$ is the optimized parameters of the neural network model after training. Consequently, the trained channel chart function can be used to estimate the UE location in a global scale coordinate frame owing to \eqref{eq:far_close_dist_loss_3}. This approach is self-supervised as it does not require labeling during the training or evaluation/test phases.
\begin{figure}[t]
    \centering
    \begin{subfigure}[b]{0.18\textwidth}
        \includegraphics[width=\textwidth]{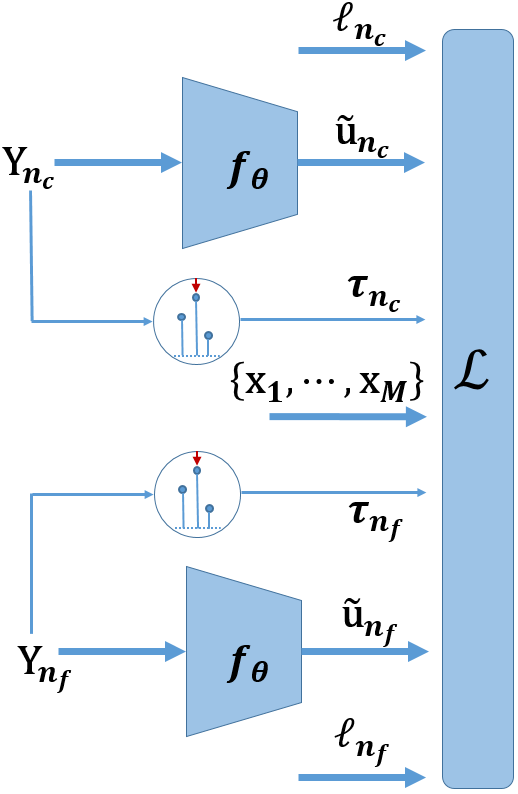}
        \caption{Training phase}
        \label{fig:diagram_train}
    \end{subfigure}\\
        \begin{subfigure}[b]{0.25\textwidth}
        \includegraphics[width=\textwidth]{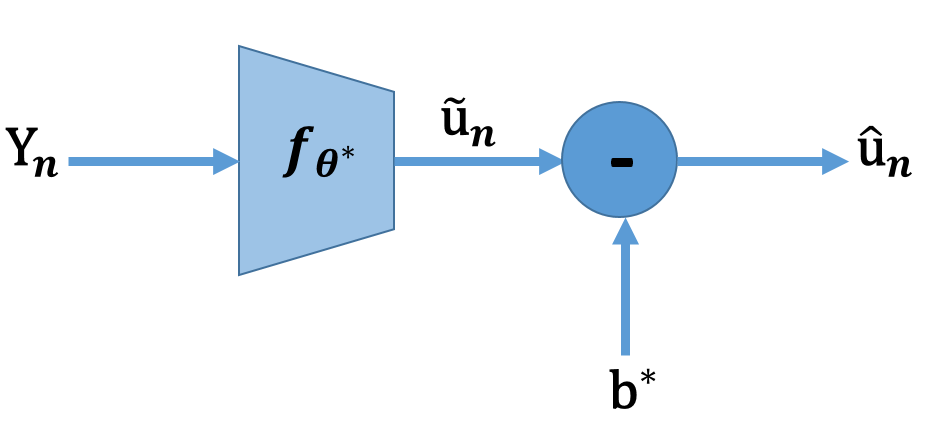}
        \caption{Test phase}
        \label{fig:diagram_train}
    \end{subfigure}
    \caption{Training and test phases of our proposed algorithm.}
    \label{fig:diagram_train_test}
\end{figure}
\subsection{Offset Estimation}
The UE location estimated using the channel chart neural network model $(f_{\boldsymbol{\theta^*}})$ trained in the last section might deviate from the ground truth by an offset. This might happen when a small set of TRPs (less than 3 TRPs) is available, which can introduce an ambiguity to the UE location estimate and result in getting stuck in a local minima. To tackle this problem, given a trained channel chart model, we formulate the following optimization problem to calculate the offset in the estimation:
\begin{equation} \label{eq:offset_estimate}
    \textbf{b}^* := \arg \min_{\textbf{b}} \sum_{n=1}^N \sum_{m=1}^M | \hat{\tau}_{m, n}(\textbf{b}) - \tau_{m, n} |, 
\end{equation}
\noindent where $\textbf{b} \in \mathbb{R}^3$ is a bias vector to be estimated, and $\hat{\tau}_{m, n}(\textbf{b})$ is the estimated ToA between the $m$-th TRP and the UE location at time step $n$ and taking into account the bias vector $\textbf{b}$. Using the trained channel chart model, $\hat{\tau}_{m, n}(\textbf{b})$ is given by:
\begin{equation}
    \hat{\tau}_{m, n}(\textbf{b}) = \frac{\|\left(f_{\boldsymbol{\theta^*}}(\textbf{Y}_{n}) - \textbf{b}\right) - \textbf{x}_m \|}{\nu}.
\end{equation}
\noindent Solving \eqref{eq:offset_estimate} will find a bias vector, minimizing the error between the estimated ToA and the measured ToA, thus improving the localization accuracy. We solve \eqref{eq:offset_estimate} using the training dataset $\mathcal{D}_{\text{tr}}$ and employing the Particle Swarm Optimization (PSO) technique \cite{488968}. We denote the optimized bias vector by $\textbf{b}^*$. 

Finally, the UE location estimate at time step $n$ by taking into account the bias vector is given by
\begin{equation}
    \hat{\textbf{u}}_n = f_{\boldsymbol{\theta^*}}(\textbf{Y}_{n}) - \textbf{b}^*.
\end{equation}
\noindent Note that $(\textbf{Y}_{n},\,\textbf{b}^*)$ are sufficient to estimate the UE location during the test time. The laser scanner data and TRPs' locations are only used during the training phase. The overall procedure of the proposed channel charting algorithm during training and test phases is illustrated in Fig. \ref{fig:diagram_train_test}.
\section{Evaluations} \label{sec: Eval}
To evaluate our proposed method against existing state-of-the-art approaches, the following benchmarks were employed:

\begin{itemize}
    \item \textbf{Classical PSO}: We compared the results with an extension of the classical TDoA-based localization algorithm in \cite{1201741}, where the PSO technique \cite{488968} is instead used to triangulate and estimate the UE location given the TDoA measurements from the surrounding TRPs.
    \item \textbf{Siamese Neural Network} \cite{10070385}: Uses pairs of CSI measurements and their corresponding Euclidean distances as a dissimilarity metric, preserved in a 2-D encoded latent space. This technique is semi-supervised since the estimated latent space requires a linear transformation to the global coordinate frame using a subset of CSI data labeled with true UE locations.
    
    \item \textbf{Triplet-based Neural Network} \cite{9448128}: It encodes triplets of CSI into a 2-D latent space and similar to Siamese, is semi-supervised and utilizes some labeled data.
    
    \item \textbf{Triplets + Bilat.} \cite{taner2023channel}: Employs a self-supervised approach using known TRPs locations and their received power in a combined triplet and bilateration loss function to train a model in the global scale coordinate frame.
\end{itemize}
The performance of the channel charting models is assessed through both quantitative and qualitative methods. Quantitative key metrics include Continuity (CT) (ensuring spatial relationships are preserved in lower-dimensional space), and Trustworthiness (TW) (ensuring charted similarities reflect true proximities) as detailed in \cite{10052099}. In addition, the CDF of the localization error at the 90th percentile (CE90), indicates localization accuracy and the spread of errors across locations. Qualitatively, evaluating the charted space through visual inspections against expected wireless environment geometries helps verify if the model accurately identifies meaningful patterns and relationships within the data.

\section{Simulation and Results}
 \begin{figure}[t]
    \centering
        \includegraphics[width=0.38\textwidth]{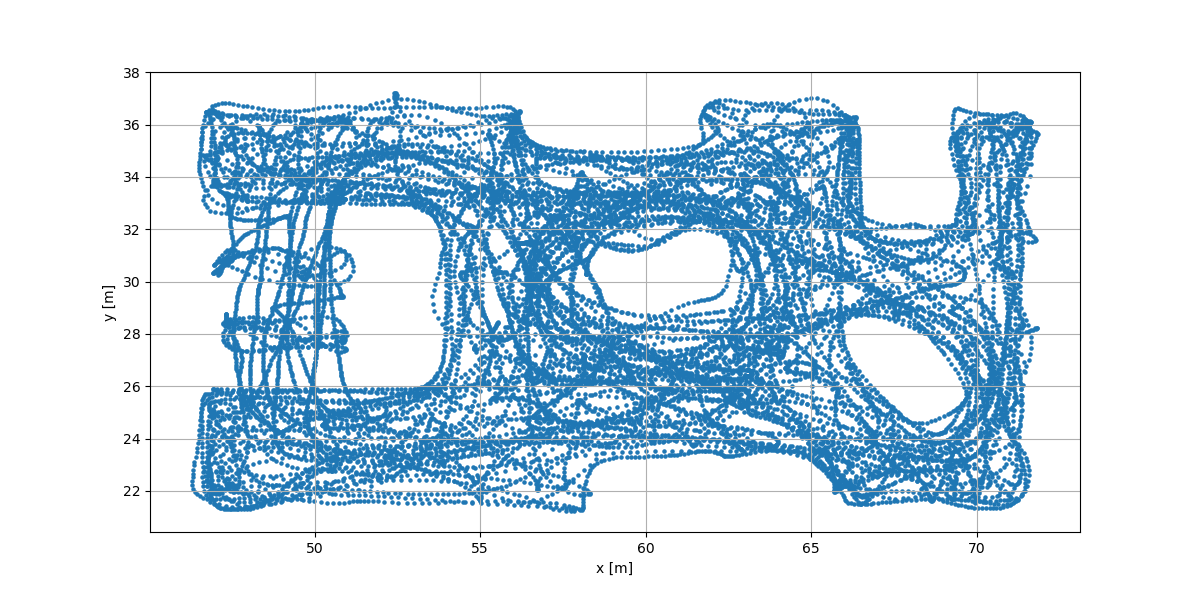}
    \caption{True trajectory during the training phase.}
    \label{fig:training_trajectory}
\end{figure}

\begin{figure*}
    \centering
    \begin{subfigure}[b]{0.18\textwidth}
        \includegraphics[width=\textwidth]{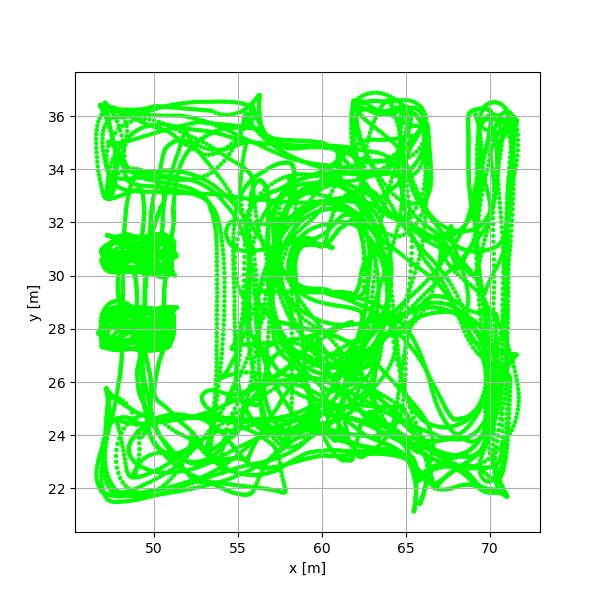}
        \caption{Ground Truth}
        \label{fig:real_pos}
    \end{subfigure}
    \begin{subfigure}[b]{0.18\textwidth}
        \includegraphics[width=\textwidth, height=\textwidth]{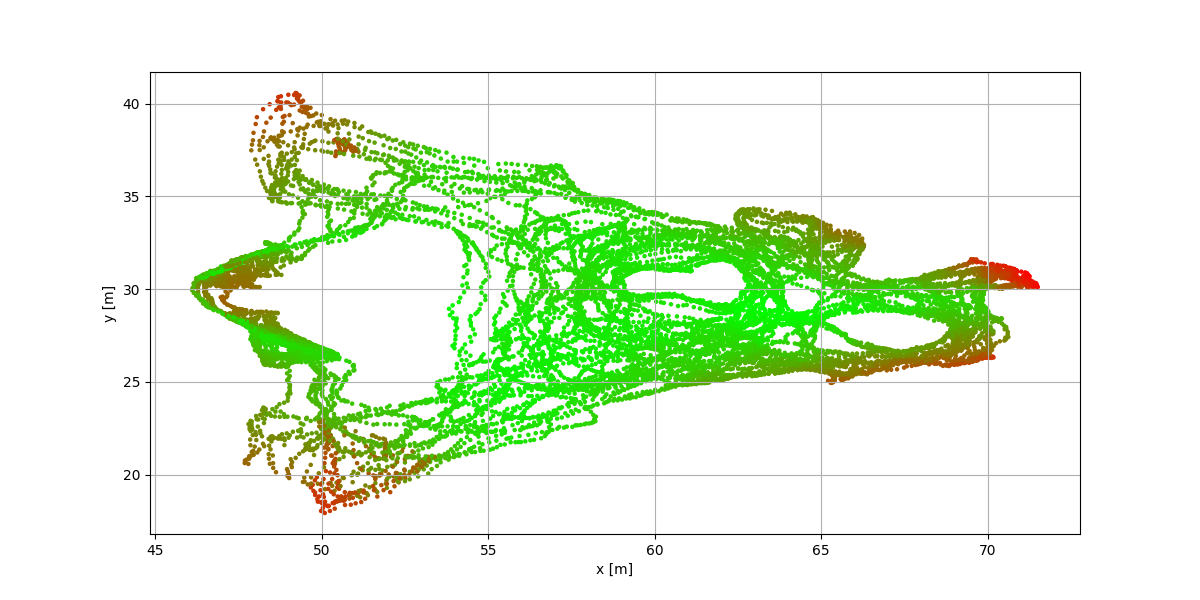}
        \caption{Siamese \cite{10070385}}
        \label{fig:pred_pos_siam}
    \end{subfigure}
    % Add a blank line for spacing if necessary
    \begin{subfigure}[b]{0.18\textwidth}
        \includegraphics[width=\textwidth, height=\textwidth]{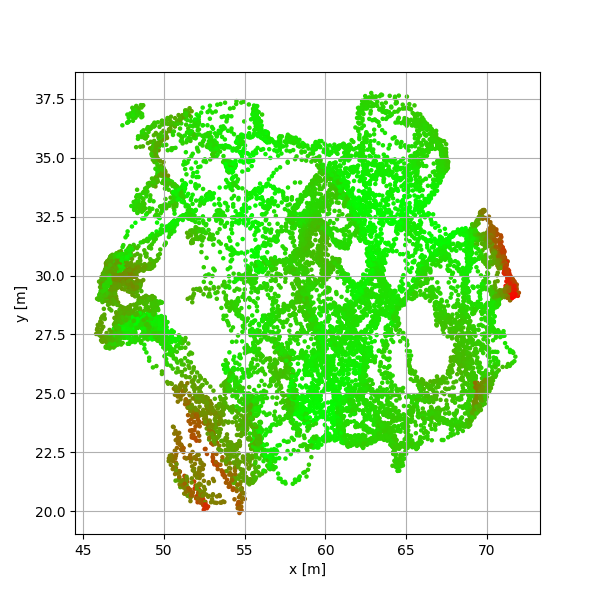}
        \caption{ Triplets \cite{9448128}}
        \label{fig:pred_pos_trip}
    \end{subfigure}
    \begin{subfigure}[b]{0.18\textwidth}
        \includegraphics[width=\textwidth, height=\textwidth]{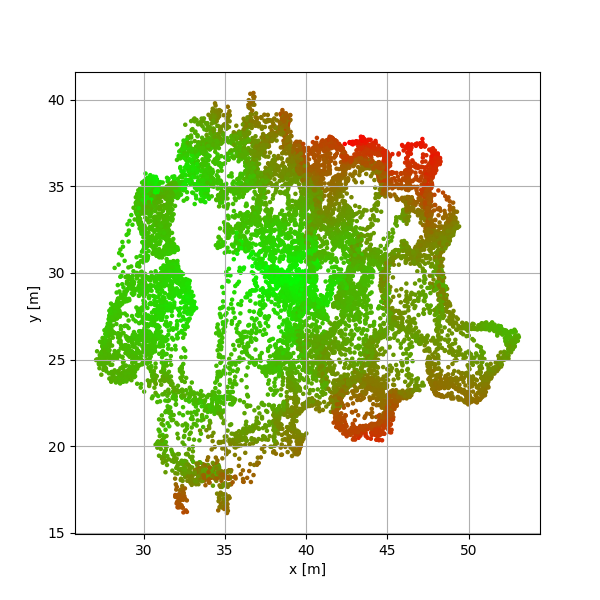}
        \caption{Trip.+Bilat. \cite{taner2023channel}}
        \label{fig:pred_pos_tb}
    \end{subfigure}
    \begin{subfigure}[b]{0.21\textwidth}
        \includegraphics[width=\textwidth]{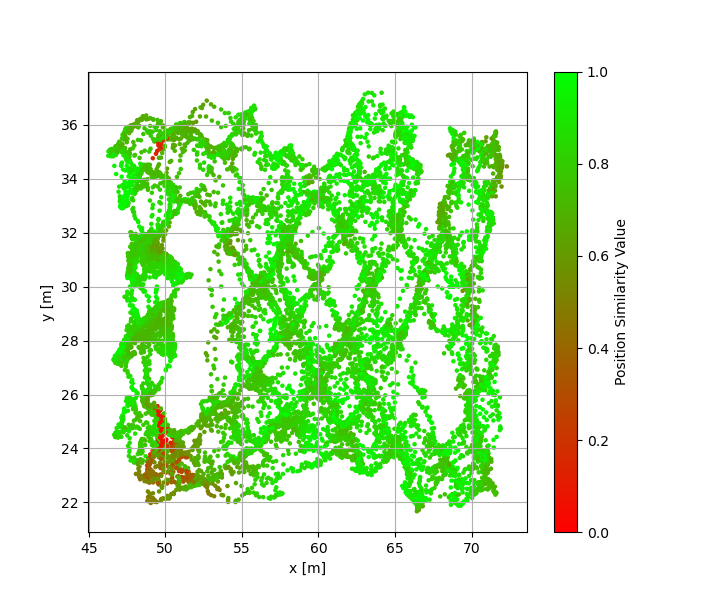}
        \caption{Ours}
        \label{fig:pred_pos_ours}
    \end{subfigure}
    \begin{subfigure}[b]{0.18\textwidth}
        \includegraphics[width=\textwidth]{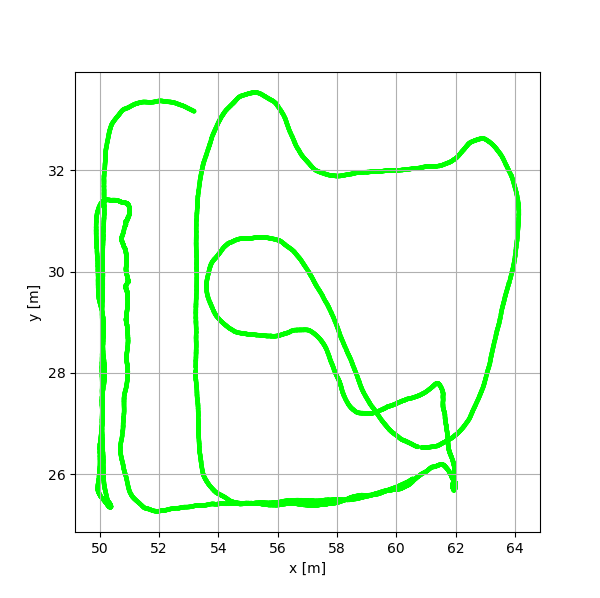}
        \caption{Ground Truth}
        \label{fig:real_pos_2}
    \end{subfigure}
    \begin{subfigure}[b]{0.18\textwidth}
        \includegraphics[width=\textwidth, height=\textwidth]{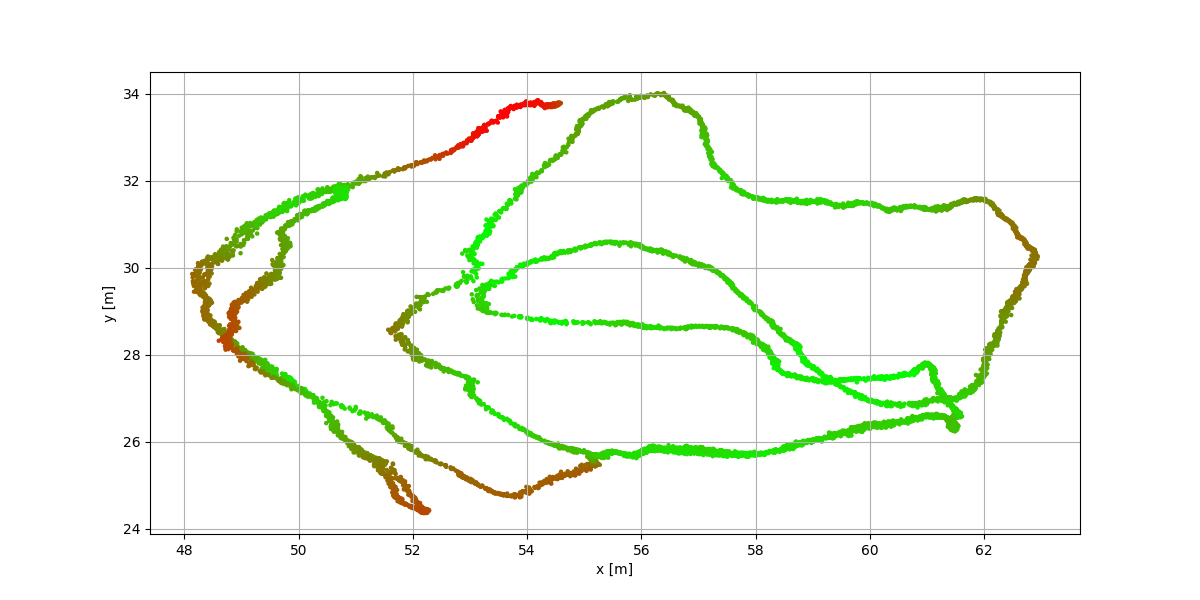}
        \caption{Siamese \cite{10070385}}
        \label{fig:pred_pos_siam_2}
    \end{subfigure}
    % Add a blank line for spacing if necessary
    \begin{subfigure}[b]{0.18\textwidth}
        \includegraphics[width=\textwidth, height=\textwidth]{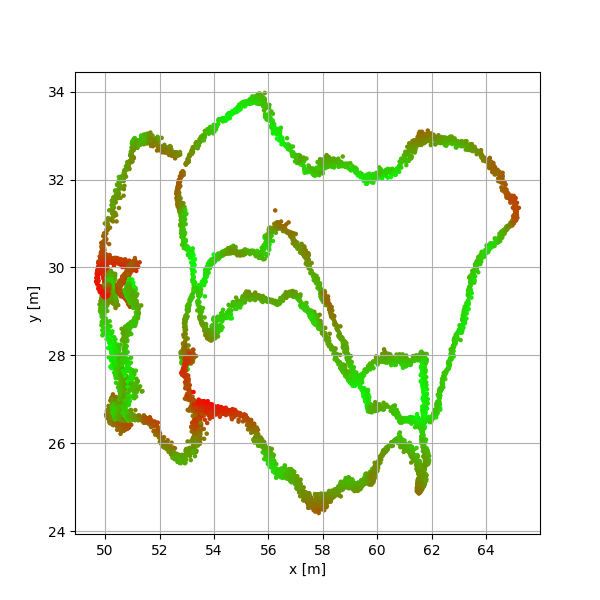}
        \caption{ Triplets \cite{9448128}}
        \label{fig:pred_pos_trip_2}
    \end{subfigure}
    \begin{subfigure}[b]{0.18\textwidth}
        \includegraphics[width=\textwidth, height=\textwidth]{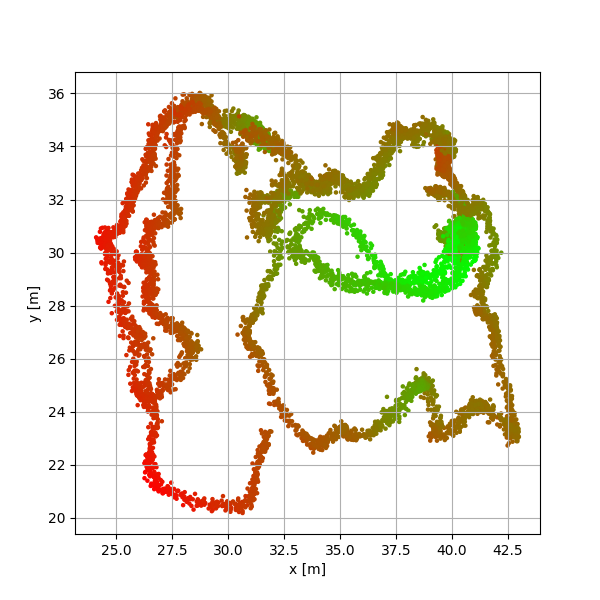}
        \caption{Trip.+Bilat. \cite{taner2023channel}}
        \label{fig:pred_pos_tb_2}
    \end{subfigure}
    \begin{subfigure}[b]{0.21\textwidth}
        \includegraphics[width=\textwidth]{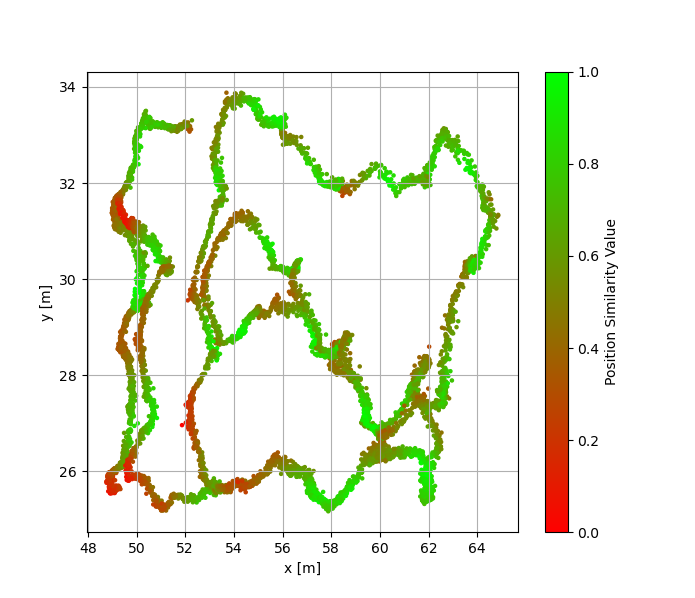}
        \caption{Ours}
        \label{fig:pred_pos_ours_2}
    \end{subfigure}
    \caption{Figures (a) to (e) are the results for test Dataset 1, and figures (f) to (j) are the results for test Dataset 2.}
    \label{fig:comparative_positions}
\end{figure*}
We consider an indoor factory scenario where there are $M=2$ LoS TRPs with a fixed height of $8 \, \text{m}$. For the sake of comparison, the training and testing trajectories are similar to the datasets from \cite{10070385} and IPIN competition 2023 Track 7, both recorded in the Fraunhofer IIS L.I.N.K. hall, where the measurements are taken in an approximately $20\times 15 \, \text{m}^2$ area. Nevertheless, we reconstructed the CIR data from the above trajectories in a Matlab environment using 3D map-based channels and a Ray-Tracing toolbox to be able to simulate the laser data in the same environment. In Fig. \ref{fig:training_trajectory}, we illustrated the ground truth trajectory taken during the training phase while in Fig. \ref{fig:real_pos} and \ref{fig:real_pos_2}, we depicted the trajectories followed by the mobile UE during the testing phase. We also define the global coordinate frame as the coordinate frame used for recording ground truth trajectories. The test datasets corresponding to the test trajectories shown in Fig. \ref{fig:real_pos} and \ref{fig:real_pos_2} are denoted by Dataset $1$ and Dataset $2$, respectively. The TRPs work in sub-6 GHz frequency with a bandwidth of $100 \,\text{MHz}$ and $122.88$\,\text{MSamples/sec} sampling rate. We choose the first $49$ elements of the CIR matrix, therefore the dimension of the input to our channel charting network is as $\textbf{Y}_n \in \mathbb{R}^{2\times 49}$. The height of the UE is assumed known and equal to $1.5\, \text{m}$ and remained constant during both the training and test phases. Therefore, we estimate the UE location in the 2D space. \footnote{This is not a limitation of the proposed algorithm and the UE location estimate can be generalized to the 3D space.} We simulated a 2D laser scanner during the training phase, which provides depth measurements with a resolution of $0.6$ degree in the angular domain, and an accuracy of $5$ cm for ranging. Both depth and CIR measurements are collected every $20 \, \text{ms}$. The architecture of the neural network used for the channel chart function is summarized in Table \ref{tab:embedding_model_summary}. In addition, We selected $\lambda_{n_c, n_f} = 5 $ when $|n_c - n_f|\leq 500$, otherwise $\lambda_{n_c, n_f} = 0$.
% \begin{equation}
%     \lambda_{n_c, n_f} = \left\{ \begin{array}{rcl}
%     5 & |n_c - n_f|\leq 500 \\ 
%     0 & \text{otherwise}
%     \end{array}\right. .
% \end{equation}

In Fig. \ref{fig:pred_pos_siam} to \ref{fig:pred_pos_ours} the results of channel charts for test Dataset $1$, and in Fig. \ref{fig:pred_pos_siam_2} to \ref{fig:pred_pos_ours_2}, the results related to test Dataset $2$ for different benchmarks are shown. For better visualization, plots are color-coded in RGB values normalized between $[0, 1.0]$, whereby points with color values closer to $1.0$ indicate more accuracy or smaller error in position estimate compared to the ground truth. It is worth mentioning that the location estimate provided by our proposed algorithm is in the global coordinate frame, very close to the ground truth. This is owing to using the ToA measurements along with the TRPs locations and the laser scanner data during the training phase.

Furthermore, the value metrics introduced in Sec. \ref{sec: Eval} for the test datasets and different benchmarks are compared in Table \ref{tab:model_comparison} which clearly shows that our proposed algorithm outperforms the benchmarks. 
Also, we compared the results with the Classical PSO benchmark which is a classical TDoA-based localization method as described in Sec. \ref{sec: Eval}. It is worth mentioning that, to triangulate the UE using such traditional algorithms, at least 3 TRPs are required, as opposed to 2 TRPs used in our algorithm. Despite using 3 TRPs, the localization accuracy of the classical method is inferior to our algorithm.

In Table \ref{tab:model_comparison}, we also present the results related to an additional experiment that we conducted by running our algorithm without utilizing the laser scanner data, which is equivalent to set $\lambda_{n_c, n_f}=0, \forall n_c, n_f$. The results confirm that the incorporation of laser scanner data can significantly improve the accuracy of localization.

\begin{table}[htbp]
\centering
\captionsetup{size=small} % Smaller caption
\caption{Embedding Model Architecture Summary}
\label{tab:embedding_model_summary}
\small % Smaller font size for the table
\begin{tabularx}{\columnwidth}{@{}Xlll@{}}
\toprule
Layer & Output Dimension & Kernel Size & Activation \\ \midrule
Conv2D & (8, 2, 49) & (3, 3) & ReLU \\
Conv2D & (8, 2, 49) & (5, 5) & ReLU \\
Conv2D & (8, 2, 49) & (8, 8) & ReLU \\
Conv2D & (16, 2, 49) & (10, 10) & ReLU \\
Flatten & (1, 1568) & - & - \\
Fully Con. & (1, 200) & - & ReLU \\
Fully Con. & (1, 100) & - & - \\
Fully Con. & (1, 2) & - & - \\ \bottomrule
\end{tabularx}
\end{table}

\begin{table}[htbp]
\centering
\captionsetup{size=small} % Smaller caption
\caption{2 TRPs Comparison of Our Model with State-of-the-Art over Datasets 1 and 2}
\label{tab:model_comparison}
\small % Smaller font size for the table
\begin{tabularx}{\columnwidth}{@{}X*{6}{c}@{}}
\toprule
 & \multicolumn{2}{c}{CT} & \multicolumn{2}{c}{TW} & \multicolumn{2}{c}{CE90 [m]} \\ 
\cmidrule(lr){2-3} \cmidrule(lr){4-5} \cmidrule(lr){6-7}
Model & 1 & 2 & 1 & 2 & 1 & 2 \\ \midrule
Classical PSO 
% \cite{1201741}
% PSO Ref. \cite{1201741} 
& 0.987 & 0.978 & 0.986 & 0.984 & 1.59 & 1.45 \\
% Semi-Sup. 
Siamese \cite{10070385}
% Ref. \cite{10070385} 
& 0.996 & 0.994 & 0.994 & 0.991 & 3.08 & 2.24 \\
% Semi-Sup. 
Triplets \cite{9448128}
% Ref. \cite{9448128} 
& 0.993 & 0.994 & 0.992 & 0.994 & 2.29 & 1.35 \\ 
% Unsup. 
Triplets+Bilat. \cite{taner2023channel}
% Ref. \cite{taner2023channel} 
& 0.991 & 0.980 & 0.990 & 0.964 & 24.14 & 22.16 \\
Ours (no Laser) & 0.995 & 0.992 & 0.995 & 0.991 & 3.98 & 3.81 \\[1mm]
\textbf{Ours} & 0.998 & 0.996 & 0.997 & 0.995 & 0.94 & 0.97 \\\bottomrule
\end{tabularx}
\end{table}

\section{Conclusions and Future Work}
In this paper, we presented a new method for channel charting that makes use of ToA measurements from nearby TRPs along with their locations. In addition, we leveraged sensor fusion by incorporating laser scanner data during the training phase of the algorithm. Our algorithm is self-supervised during the training and test phases, requiring no geometrical models or user position ground truth. Simulation results demonstrated that our algorithm achieves sub-meter level localization accuracy 90$\%$ of the time, surpassing the state-of-the-art channel charting techniques and the traditional triangulation-based approaches.

Channel charting leveraging data fusion is still in its infancy and requires further investigation. For future research, we will extend this work by exploring NLoS scenarios as well as exploiting additional sources of measurements, such as AoA. Moreover, $\lambda_{n_c, n_f}$ can be selected more dynamically.

\section*{Acknowledgment}
Part of the work by O. Esrafilian was funded via HUAWEI France supported Chair on Future Wireless Networks at EURECOM. In addition, part of this work has been supported by the ”France 2030” investment program through the projects 5G-OPERA and GEO-5G.
\bibliographystyle{IEEEtran}
\bibliography{IEEEabrv,ref}
\vspace{12pt}
\end{document}